\documentclass[11pt,preprint]{aastex}

\newcommand{\um}{$\mu$m}
\newcommand{\bump}{1.6$\mu$m bump}

\newcommand{\jh}{$J$$-$$H$}
\newcommand{\hk}{$H$$-$$K$}
\newcommand{\jk}{$J$$-$$K$}
\newcommand{\kl}{$K$$-$$L$}
\newcommand{\hl}{$H$$-$$L$}
\newcommand{\jl}{$J$$-$$L$}

\newcommand{\iracfilters}{3.6\um, 4.5\um, 5.8\um, 8\um}
\newcommand{\iracL}{m$_{3.6}$$-$m$_{8}$}
\newcommand{\iracM}{m$_{4.5}$$-$m$_{8}$}
\newcommand{\iracS}{m$_{5.8}$$-$m$_{8}$}
\newcommand{\lap}{$\lesssim$}
\newcommand{\gap}{$\gtrsim$}
\newcommand{\ebv}{$E(B-V)$}
\newcommand{\Zsun}{ $Z_{\odot}$}
\newcommand{\agein}{age$_{in}$}
\newcommand{\agefit}{age$_{fit}$}
\newcommand{\zin}{$z_{in}$}
\newcommand{\zfit}{$z_{fit}$}
\newcommand{\za}{$z$$\approx$}

\slugcomment{AJ, in press}

\shorttitle{The 1.6\um\ Bump}
\shortauthors{Sawicki}

\begin{document}


\title{The 1.6\um\ Bump as a Photometric Redshift Indicator}

\author{Marcin Sawicki} 
\affil{
Dominion Astrophysical Observatory, 
Herzberg Institute of Astrophysics,
National Research Council, 
5071~West Saanich Road, 
Victoria, B.C., V9E 2E7, 
Canada
}
\email{marcin.sawicki@nrc.ca}

\begin{abstract}
I describe the principle of using the 1.6\um\ H$^-$ spectral feature
as a photometric redshift indicator and demonstrate that the technique
holds promise by successfully recovering the redshifts of a small
sample of $z$=0--1 galaxies using only their infrared ($JHKL$)
photometry.  I then consider the applicability of the technique to the
3.6--8\um\ SIRTF filter set and investigate the systematic errors that
could arise in photometric redshifts from random photometric errors or
from a mismatch between target galaxies and fitting templates in
metallicity, star formation history, and amount of interstellar dust.
It appears that SIRTF near-IR data alone should be sufficient to
estimate redshift of most galaxies that are at $z$\gap 1.5 and are
dominated by stellar populations older than \gap 20Myr.  Galaxies
whose photometric fits indicate them to be at lower redshifts,
\zfit\lap1.5, or dominated by very young stellar populations,
\agefit\lap20Myr, suffer from severe degeneracies in photometric
redshift, and a reliable photometric determination of their redshifts
will have to include either IR observations at shorter wavelengths
($H$ and $K$) or optical data.  Overall, it appears that --- with care
and caveats --- the 1.6\um\ bump can provide a powerful way of
estimating redshifts of distant galaxies in deep infrared imaging
surveys that will soon be provided by SIRTF, and, eventually, by NGST.
\end{abstract}


\section{INTRODUCTION}

The study of galaxy formation and evolution will gain an enormous
boost from space-borne observations in the infrared (IR), first with
the advent of the Space Infrared Telescope Facility (SIRTF) and then,
at the end of this decade, with the appearance of the Next Generation
Space Telescope (NGST).  SIRTF, scheduled for launch in early 2003,
will allow us to efficiently observe distant galaxies at mid- and
far-infrared wavelengths.  In particular, SIRTF's Infrared Array
Camera (IRAC) will operate at 3.6\um, 4.5\um, 5.8\um, and 8\um,
providing a window on the rest-frame near-IR properties of galaxies at
$z$\gap1.  A number of deep SIRTF imaging surveys that will detect
many thousands of galaxies to high redshift are already scheduled
(e.g., GOODS\footnote{see http://www.stsci.edu/science/goods/},
SWIRE\footnote{see http://www.ipac.caltech.edu/SWIRE/}), and it will
be enormously important to efficiently estimate redshifts of galaxies
in these samples.  However, large-scale spectroscopic follow-up of
these IR-selected surveys will be very time-consuming and only
small sub-samples of objects are likely to be targeted.  Instead,
spectroscopic redshifts are likely to be complemented by the less
precise but much more efficient photometric redshifts.

Photometric redshifts enjoy great popularity as a way to estimate the
redshifts of faint galaxies (e.g., Connolly et al.\ 1995; Gwyn \&
Hartwick 1996; Sawicki, Lin, \& Yee 1997; Fern\'andez-Soto, Lanzetta,
\& Yahil 1999; Rudnick et al.\ 2001).  Traditionally, photometric
redshifts rely on the photometric signatures of the Lyman and 4000\AA\
breaks to estimate redshifts from optical and near-IR data.  However,
the availability of mid-IR data from SIRTF (and, later, NGST) surveys
opens up the possibility of using a hitherto untried photometric
redshift technique that could obviate the need for optical imaging of
SIRTF and NGST survey fields, or provide independent redshift
estimates where such imaging is already present.  Specifically, the
spectral ``bump'' at 1.6\um, caused by the minimum in the opacity of
the H$^-$ ion present in the atmospheres of cool stars, is a nearly
ubiquitous feature of almost all stellar populations and the
photometric signature of this bump can be expected to provide a means
of estimating redshifts to distant galaxies.

The importance of the imprint of the \bump\ on SIRTF IRAC colors has
been appreciated for a long time (e.g., Wright, Eisenhardt, \& Fazio
1994), and indeed the filter transmission curves of the IRAC filters
have been designed partially with photometric redshifts in mind by
modeling the expected colors of high-$z$ galaxies (Simpson \&
Eisenhardt 1999).  The purpose of the present paper is to demonstrate
the potential of the 1.6\um\ bump as a photometric redshift indicator
and to explore some of the possible limitations of the technique at
high redshift.  Section~\ref{lowz} illustrates, with real data at
0$<$$z$$<$1, that the 1.6\um\ bump {\it is} an effective photometric
redshift indicator and describes the source of the photometric
redshift signal.  Section~\ref{sirtf} discusses the applicability of
the technique to SIRTF IRAC data and explores some of the limitations
that could be present at high redshift due to effects such as
metallicity, dust, and star formation history.  The main conclusions
of the paper are summarized and discussed in Section~\ref{summary}.

\section{PRINCIPLE AND PROOF OF CONCEPT}\label{lowz}

\subsection{The 1.6\um\ Bump}

The opacity of the H$^-$ ion has a minimum at $\sim$1.6\um\ (e.g.,
John 1988), which imprints itself as a maximum, or ``bump,'' on the
spectral energy distributions (SEDs) of cool stars.  Consequently, the
SEDs of composite stellar populations that contain significant numbers
of such stars can be expected to possess the same spectral imprint.
This assertion is illustrated in Fig.~\ref{seds.fig}, which shows the
spectral energy distributions of model stellar populations from the
1996 version of the GISSEL spectral synthesis code of Bruzual \&
Charlot (1993).  The 1.6\um\ bump is easily discernible in all but the
extremely young ($\sim$1~Myr) stellar populations, where it is swamped
by the essentially power-law spectra of very hot, massive young stars.
Given the strength and near universality of the 1.6\um\ bump, it is
reasonable to expect that it will imprint a significant signal on the
colors of distant galaxies, and hence, as was pointed out by Simpson
\& Eisenhardt (1999), that it might make a good photometric redshift
indicator.

\subsection{The 1.6\um\ Bump in the Hubble Deep Field}\label{HDF}

While the 1.6\um\ bump has been suggested as a photometric redshift
indicator before, its feasibility remains to be tested with real data.
The $JHKL$ filter set straddles the \bump\ at $z$$\approx$0--1, and so
$JHKL$ photometry of Hubble Deep Field (HDF; Williams et al.\ 1996)
galaxies with spectroscopic redshifts will be used to test if the
\bump\ can provide a strong enough photometric imprint to constrain
redshifts.  Note that the $JHKL$ filter set provides a reasonable
surrogate of the IRAC quartet of bandpasses, albeit shifted by a
factor of $\sim$2.5--3 in observed wavelength: while filters of the
$JHKL$ set cover the rest-frame \bump\ for galaxies at \za0--1, the
corresponding redshift range for the IRAC filters is \za1.8--4.
Consequently, $JHKL$ photometric redshifts of $z$$<$1 galaxies provide
a useful demonstration of how well SIRTF may do at these higher
redshifts.

Nine galaxies were detected at $L$-band (3.2\um) in the deep Keck
imaging of the HDF by Hogg et al.\ (2000).  All nine have redshifts at
0$<$$z$$<$1, as measured by Cohen et al.\ (1996, 1999) and Lowenthal
et al.\ (1997).  The $L$-band photometry of these objects was adapted
from Hogg et al.\ (2000), while their $JHK$ magnitudes were measured
from the public Kitt Peak infrared images of the HDF (Dickinson, 1998;
Dickinson et al.\ 2002, in prep.).  The $JHK$ magnitudes were measured
in 2\arcsec apertures using the SExtractor package (Bertin \& Arnouts,
1996), while the $L$-band magnitudes of Hogg et al.\ (which they
measured in 2\arcsec apertures but then extrapolated to 6\arcsec\
ones) were transformed back to 2\arcsec apertures by scaling (in
reverse of the procedure used by Hogg et al.) the $L$-band 6\arcsec\
magnitudes by the ratio of $H$-band fluxes measured in 2\arcsec\ and
6\arcsec\ apertures.  The resulting 2\arcsec-aperture magnitudes are
listed in Table~\ref{photometry.tab}.

Photometric redshifts of the nine galaxies were estimated by comparing
their observed broadband spectral energy distributions (as listed in
Table~\ref{photometry.tab}) with model templates generated from the
1996 version of the GISSEL spectral synthesis package (Bruzual \&
Charlot 1993).  Solar-metallicity GISSEL spectra of constant star
formation rate (SFR), and with ages spanning the full available range
of models (0--20 Gyr), were used as the starting point in the
construction of the templates.  These model spectra were redshifted
onto a grid of redshifts spanning $z$=0--1.5 in steps of $\delta
z$=0.05, and were then integrated through the filter transmission
curves (FTCs) to produce the template broadband fluxes
\begin{equation}\label{ftpt.eq}
f_{tpt}(i) = \int SED(\lambda(1+z)) \cdot FTC(i,\lambda) d\lambda
\end{equation}
in the four filters $i$=($J,H,K,L$). The resulting grid contains 6851
templates, each of four broadband fluxes, and spans
0$\leq$$z$$\leq$1.5 linearly and 0$\leq$age$\leq$20~Gyr
quasi-logarithmically.

For each observed object, the observed fluxes were compared to each
template in the template grid by computing the goodness-of-fit measure
\begin{equation}\label{chisq.eq}
\chi^2 =\sum_{i}\left[{ \frac{f_{obs}(i)-s\cdot
f_{tpt}(i)}{\sigma(i)}}\right]^2,
\end{equation}
where $f_{obs}(i)$ and $\sigma(i)$ are the observed flux and its
uncertainty in the $i^{th}$ filter, $f_{tpt}(i)$ is the flux of the
template in that filter, and $s$ is the scaling between the observed
and template fluxes that can be computed analytically using
\begin{equation}\label{s.eq}
s = \sum_{i} \frac{f_{obs}(i)\cdot f_{tpt}(i)}{ \sigma^2(i)}
/
\sum_{i} \frac{f^2_{tpt}(i) }{ \sigma^2(i)}.
\end{equation}
For each object, the best-fitting template --- and, hence, the most
likely redshift --- is then identified as the one that produces the
smallest value of $\chi ^2$.  Note that variants of this procedure are
a common way of computing photometric redshifts using model templates
(e.g., Gwyn \& Hartwick 1996; Sawicki et al.\ 1997, Arnouts et al.\
1999), with the key difference being that the data and templates used
here do not span the traditional 4000\AA\ or Lyman breaks, but,
instead, cover the \bump.

The fidelity of the photometric redshifts obtained using {\it solely}
$JHKL$ photometry is remarkably good, as is shown in
Fig.~\ref{HDF.fig} which compares the photometric and spectroscopic
redshifts of the nine HDF galaxies (Table~\ref{photometry.tab} lists
the $z_{phot}$ values object by object.).  The scatter between
photometric and spectroscopic redshifts is only $\sigma_{\Delta
z}$=0.09, which is similar to the scatter typically obtained over the
same redshift range in the HDF using the Balmer-break sensitive
optical photometric redshift techniques (e.g., Sawicki et al.\ 1997;
Fern\'andez-Soto et al.\ 1999).  Although the comparison between
photometric and spectroscopic redshifts is based on the very small
sample of nine galaxies, the tight correlation between $z_{phot}$ and
$z_{spec}$ demonstrates, for the first time, that the 1.6\um\ bump is
a viable indicator of redshift.

\subsection{The Source of Photometric Redshift Signal}\label{source_of_signal}

To understand how the 1.6\um\ bump produces a photometric redshift
signature, it is instructive to consider how the infrared colors of
galaxies depend on redshift.  The $JHKL$ filter set used in
\S\ref{HDF} can be represented as six constituent colors;
Fig.~\ref{JHKL_colours.fig} shows these six colors for GISSEL models
of constantly star-forming stellar populations of seven different
ages.  The solid lines represent colors of stellar populations ranging
in age from 20 Myr to 20 Gyr, while the dotted line is for the
extremely young 1 Myr-old population.

\begin{figure}
\plotone{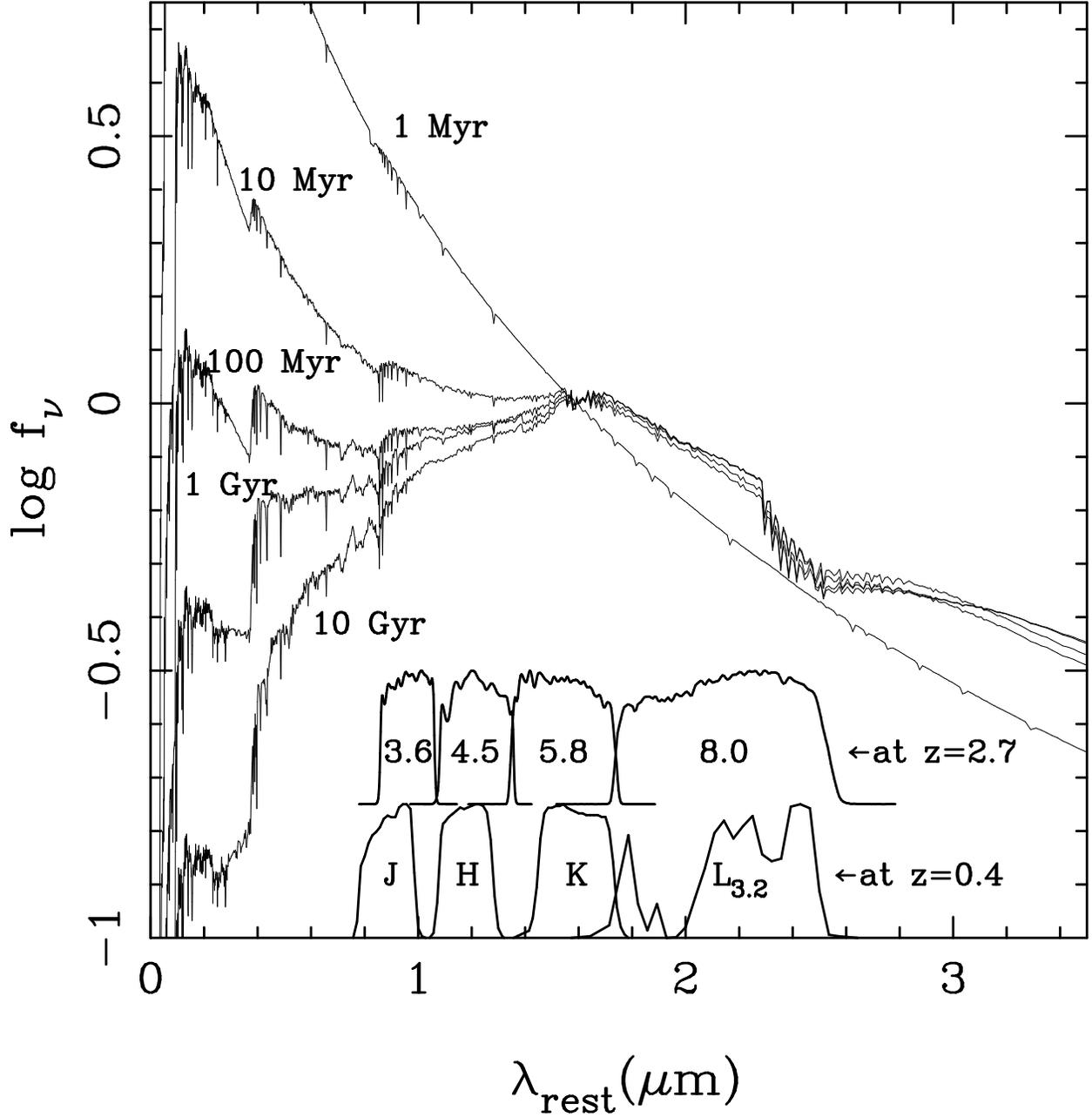}
\caption{\label{seds.fig} Model spectral energy distributions from the
1996 version of the GISSEL library of Bruzual \& Charlot (1993).
Shown are SEDs of solar metallicity stellar populations that are
forming stars at a constant rate with a Salpeter IMF. The SEDs are
normalized at 1.6\um.  The 1.6\um\ bump is a prominent feature in all
but the very youngest stellar populations.  At the bottom of the plot
are shown the filter transmission curves of the SIRTF IRAC filters
(redshifted to show them at $z$=2.7 in the rest-frame of the SEDs) and
of the $JHKL$ filter set (here redshifted to $z$=0.4) that was used in
fitting the nine HDF galaxies in Sec.~\ref{HDF}.}
\end{figure}

\begin{figure}
\plotone{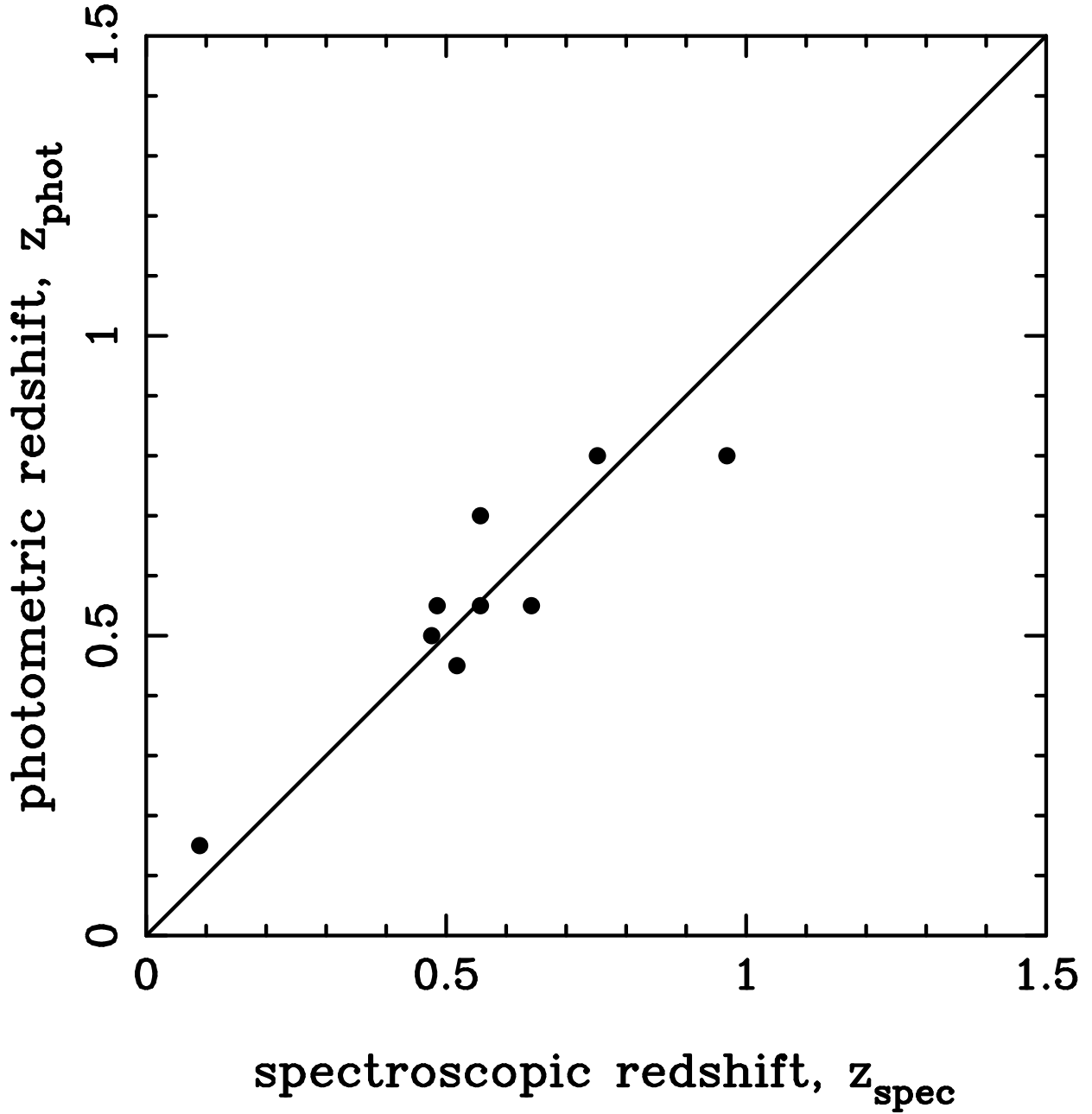}
\caption{\label{HDF.fig} Comparison of photometric and spectroscopic
redshifts for galaxies in the Hubble Deep Field that have $JHKL$
photometry.  The photometric redshifts were derived using only the
$JHKL$ data and so rely solely on the \bump.  The scatter between
photometric and spectroscopic redshifts is small ($\sigma_{\Delta z}$
= 0.09) demonstrating that the 1.6\um\ bump is a viable redshift
indicator.}
\end{figure}

\begin{figure}
\plotone{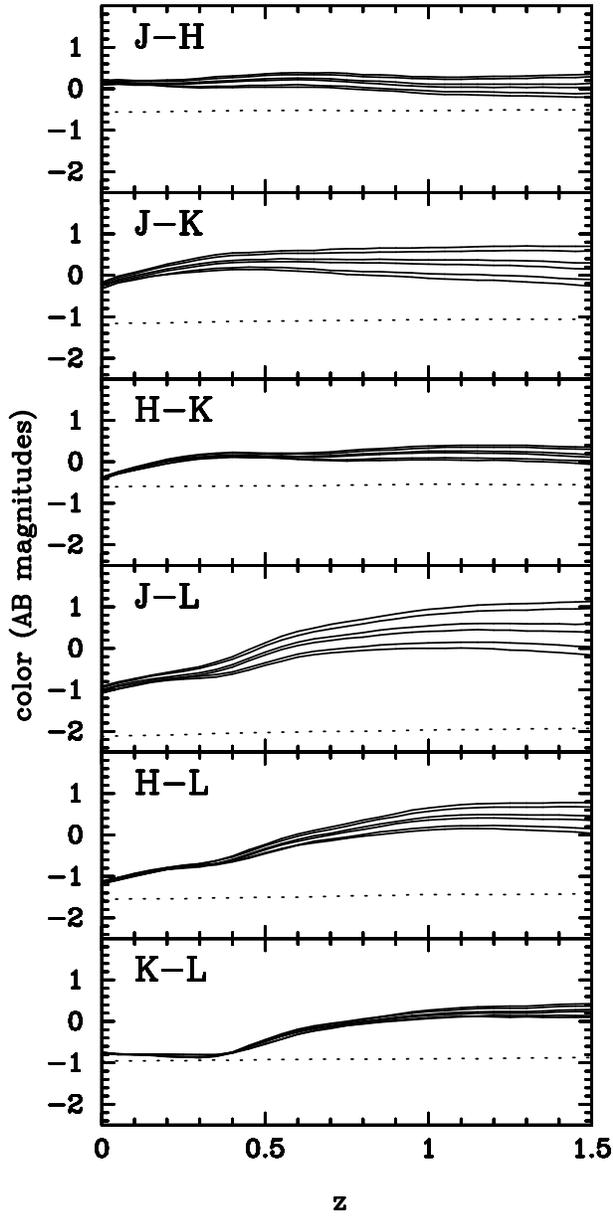}
\caption{\label{JHKL_colours.fig} Dependence of infrared galaxy colors
on redshift.  Colors were computed from solar metallicity, constant
star formation rate GISSEL models.  Within each panel, the solid lines
are, from top to bottom, for model ages of 20~Gyr, 10~Gyr, 2 Gyr, 1
Gyr, 100 Myr, and 20 Myr, and the dotted line is for the extremely
young 1 Myr-old model.}
\end{figure}

For the extremely young, 1 Myr-old model, all six colors are virtually
constant with redshift: the spectra of the very young, hot stars that
dominate this ultra-young stellar population are essentially
featureless Rayleigh-Jeans power laws (see Fig.~\ref{seds.fig}) and
because redshifting a power law does not affect its spectral shape,
the colors of extremely young stellar populations remain degenerate
with redshift.  Consequently, the redshift of a galaxy dominated by an
extremely young stellar population cannot be determined
photometrically.  Fortunately, stellar populations older than a few
Myr develop the 1.6\um\ bump (see Fig.~\ref{seds.fig}), and it is the
passage of this bump through the $JHKL$ filter set that imprints
itself on the colors of galaxies in Fig.~\ref{JHKL_colours.fig}.  For
the remainder of this section, extremely young ($\sim$1 Myr-old)
stellar populations will be ignored and the analysis will focus on the
more prosaic models with age$\geq$10 Myr.

Clearly, colors which are degenerate with redshift cannot provide
redshift information: for example, the \jh\ colors of all model ages
are essentially flat with redshift and so \jh\ contains little
information about redshift.  In contrast, where a color changes
steeply with redshift, such as \kl\ over
0.4$\lesssim$$z$$\lesssim$0.8, it provides strong redshift signal.
However, to be useful, the color must not only change steeply, but
must also be unique with redshift: thus, although the \jl\ color
changes steeply with redshift at \jl=$-$0.5, it does not constrain
redshift well since models of different age produce that same color at
different redshifts.  Therefore, the requirements for a color to be a
good photometric redshift indicator are that (1) it change steeply
with redshift and (2) it be unique for a given redshift.

Fig.~\ref{JHKL_colours.fig} shows that \jh\ is a poor redshift
indicator over all redshifts $z$$<$1.5; \hk\ is fine only at
$z$$\sim$0; \jk\ would be fine at low redshift were it not for the
fact that in this color $z$$\approx$0 objects are degenerate with
young stellar populations at $z$$\gtrsim$1; \jl\ is fine at
$z$$\approx$0, \hl\ at 0\lap $z$\lap 0.4, and \kl\ at 0.4\lap $z$\lap
0.8.  Thus, the success of the $JHKL$ filter set at determining the
redshifts of the nine HDF galaxies is primarily due to the \hl\ color
below $z$$\approx$0.4 and \kl\ at higher $z$.

The two most useful colors, \hl\ at $z$$\lesssim$0.4 and \kl\ at
$z$\gap 0.4, have an important feature in common: they both contain
the longest-wavelength $L$ filter.  The role of this filter may be
appreciated in Fig.~\ref{seds.fig}: At all redshifts $z$\lap 1, the
$L$ filter is redward of the \bump\ --- i.e., at wavelengths at which
the spectra of virtually all stellar populations are nearly identical.
Thus, the $L$ filter provides a universal ``anchor,'' which is
independent of galaxy type. The role that the other two filters ($H$
and $K$) play now becomes apparent: while $L$ acts as a
redshift-invariant anchor, the \bump\ moves with increasing redshift
first through $J$ and then through $K$.  The passage of the \bump\
through the $J$ and $K$ bandpasses depresses the fluxes in these
filters with respect to the invariant $L$-band anchor, thus resulting
in the steep yet age-invariant evolution of the \hl\ and \kl\ colors
over 0\lap $z$\lap 0.4 and 0.4\lap $z$\lap 0.8, respectively.

As is shown in Fig.~\ref{seds.fig}, spectra of galaxies are virtually
identical redward of rest-frame 1.5\um\ and so the \hl\ and \kl\
colors of galaxies are independent of age over the, respectively,
0\lap$z$\lap0.4 and 0.4\lap$z$\lap0.8 redshift ranges.  At wavelengths
below rest-frame 1.5\um, galaxy colors become strongly dependent on
age and, consequently, at $z$\gap 0.4, the \hl\ color develops an
age-redshift degeneracy.  However, the \kl\ color takes over at
$z$$\gtrsim$0.4 as a redshift indicator, and the \hl\ color now can
assume a different role --- that of providing information about the
evolutionary state of the galaxy's stellar population!

The above discussion illustrates the general principle of selecting
filters for measuring redshifts using the \bump: One needs an
``anchor'' filter which will at all redshifts of interest remain in
the age-invariant part of the spectrum redward of the \bump.
Furthermore, one needs a filter whose bandpass corresponds to the
wavelengths through which the \bump\ will be passing in the redshift
range of interest.  A third filter can then be deployed in the
age-sensitive region blueward of the \bump\ to gain information about
the state (age) of the galaxy's stellar population.  Finally, to cover
an extensive redshift range, a set of adjacent filters is necessary so
that the filters can take over from each other at different redshifts.
In practice, of course, fluxes from all available filters contribute
when the photometric redshift is determined via $\chi^2$ minimization
of broadband spectral energy distributions; nevertheless, the bulk of
the photometric redshift signal at a given redshift $z$ comes from the
combination of a long-wavelength anchor bandpass and a bump-containing
filter at $\lambda_{filter}$$\approx$$(1+z)$1.6\um.

\section{THE 1.6\um\ BUMP AND SIRTF}\label{sirtf}

Armed with the understanding of how the \bump\ interacts with
broadband filters, let us consider how the SIRTF IRAC \iracfilters\
bandpasses can be used to estimate redshifts.  The 8\um\ bandpass is
the reddest of the IRAC filters and so it naturally serves as the
long-wavelength anchor.  In combination with the 8\um\ anchor, the
bluer 3.6\um, 4.5\um, and 5.8\um\ bandpasses then provide the bulk of
the photometric redshift signal over different redshift ranges.  This
principle is illustrated in Fig.~\ref{IRAC_colours.fig}, which shows
the IRAC colors of model stellar populations (see also Fig.9 of
Simpson \& Eisenhardt 1999).  The \iracL\ color provides a good
photometric redshift indicator over 1.3\lap$z$\lap2, while becoming
degenerate to galaxy age at higher redshifts; the \iracM\ color works
at 2\lap$z$\lap2.7, while \iracS\ is good at 2.7\lap$z$\lap3.

Note that at low redshift ($z$\lap1--2, depending on the central
wavelength of the blue filter) IRAC filters are not probing the region
around the \bump.  Instead, they are sensitive to the presence of the
CO absorption band at 2.4\um.  The interplay of the IRAC filters with
the CO band results in a series of ``wiggles'' in the colors at
$z$\lap1--2.  These wiggles make the IRAC colors degenerate at low
redshift and so restrict the redshift range over which IRAC photometry
can be used to measure redshift.

Let us now consider in more detail what biases and uncertainties can
be expected when trying to determine redshifts using SIRTF IRAC
photometry.  Photometric redshift measurements can be biased,
sometimes catastrophically, by random errors in photometry, as well
as, in a more systematic way, by incorrect assumptions about the
target galaxy's spectral energy distribution.
Section~\ref{photometricerrors} will examine the biases that can be
introduced by simple Gaussian scatter in photometric measurements,
while \S\ref{systematics} will examine the problems that may arise
from incorrect assumptions about the metallicities, star formation
histories, and dust content of high-$z$ galaxies.

\subsection{Random Photometric Errors}\label{photometricerrors}

In practice, photometric redshifts of SIRTF-selected galaxies will be
computed using all four IRAC bandpasses (plus possibly additional
observations at other wavelengths), using a method similar to the
template-fitting technique that was used to determine photometric
redshifts of HDF $z$\lap1 galaxies in \S\ref{HDF}.  To understand how
random Gaussian photometric errors can affect the measurement of
photometric redshifts, it is instructive to consider how such
photometric errors propagate through eq.(\ref{chisq.eq}).

To simulate the IRAC colors of high-$z$ galaxies, a grid of models was
computed based on the constant star formation GISSEL spectra covering
the full range of available ages 0$\leq$age$\leq$20Gyr and spanning
0$\leq$$z$$\leq$5 in steps of $\delta z$=0.05.  Since high-$z$
galaxies can be expected to have sub-solar metallicities,
$Z$=0.2\Zsun\ GISSEL SEDs were used.  These SEDs were integrated
through the four SIRTF IRAC filter transmission curves (see
eq.[\ref{ftpt.eq}]) to produce a grid of predicted fluxes.  Sample
``observed'' galaxies were taken from this model grid and were fit via
eq.(\ref{chisq.eq}) with the full grid of models as the template grid.
Each ``observed'' galaxy was assigned photometric errors of
$\sigma(i)$=(0.05, 0.05, 0.75, 0.1) in the four IRAC filters,
$i$=(\iracfilters), which correspond to the errors expected for a
$z$=3 $L^*$ galaxy in the GOODS fields.

Fig.~\ref{simulate_auto.fig} shows the fit results in redshift-age
space for ten ``observed galaxies'' --- at two input, i.e. ``true,''
ages (\agein=50Myr in the left-hand panels and 1Gyr in the right-hand
panels) and five redshifts, \zin=0.5, 1, 2, 3, and 4 (top to bottom).
In each panel, the true redshift-age location of the ``observed''
galaxy is shown as a filled circle; since the ``observed'' galaxies
are taken directly from the fitting grid, this is also, by
construction, the location of the best-fit template identified by the
fitting process.  Regions of fit confidence are identified by
selecting those templates for which $\chi ^2$ values depart by less
than a certain amount from the best-fitting $\chi_{min}^2$; shown are
regions of 95\% confidence for ``observed'' data with random Gaussian
photometric errors of $\sigma(i)$=(0.05, 0.05, 0.075, 0.1) in the IRAC
filters $i$=(\iracfilters).  It is the shapes of these confidence
regions that illustrate the biases and scatter that will exist in
photometric redshifts derived from SIRTF IRAC data.

\begin{figure}
\plotone{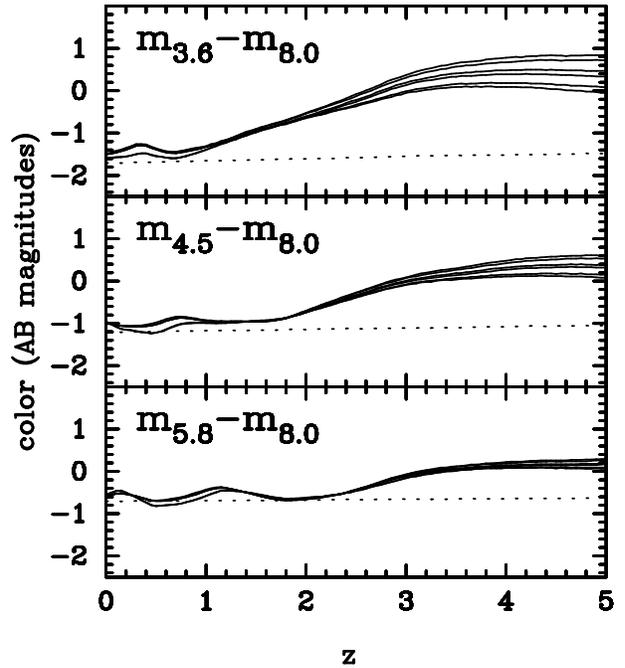}
\caption{\label{IRAC_colours.fig} Redshift dependence of galaxy colors
in the SIRTF IRAC bandpasses.  As in Fig.~\ref{JHKL_colours.fig},
solid lines show colors of solar metallicity, constant star formation
rate GISSEL spectral synthesis models spanning ages of 20 Myr -- 20
Gyr, while the dotted line is for the extremely young, 1 Myr-old
model.}
\end{figure}

\begin{figure}
\plotone{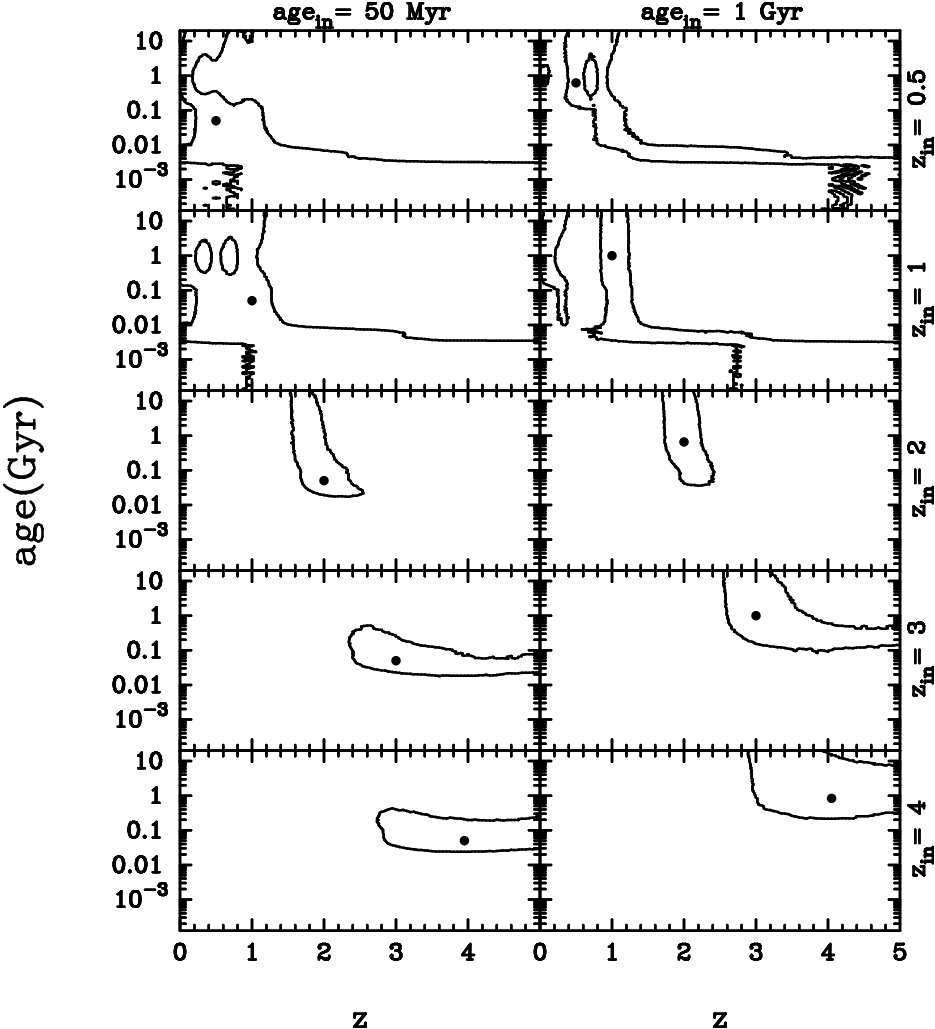} 
\caption{\label{simulate_auto.fig} Confidence in recovery of redshift
from SIRTF IRAC photometry.  Constantly star-forming galaxies of 20\%
solar metallicity, and with ages of 50 Myr (left-hand panels) and 1
Gyr (right-hand panels) and located at $z$=0.5, 1, 2, 3, 4 (from top
to bottom), were fitted with a grid of templates with the same
metallicity and star formation history.  Filled circles indicate the
best fit solutions (\zfit, \agefit) which are, by construction,
identical to the true, input values of redshift and age (\zin,
\agein).  The contours enclose regions of parameter space that will
contain 95\% of objects which have (0.05, 0.05, 0.075, 0.1) mag
uncertainties in the four IRAC filters.  }
\end{figure}

The inspection of Fig.~\ref{simulate_auto.fig} leads to a number of
observations:

\begin{enumerate} 
\item The girth of the contours in the top two rows of
Fig.~\ref{simulate_auto.fig} is indicative of the expected scatter in
photometric redshifts at $z$\lap1.5 and illustrates that the
determination of photometric redshifts with IRAC data alone is
unfeasible at z\lap1.5.  This large scatter within 0$\leq$$z$\lap1.5
results from the degeneracies in IRAC colors seen at low $z$ in
Fig.~\ref{IRAC_colours.fig}.  Thus, IRAC colors alone are insufficient
for determining redshifts of $z$\lap1.5 galaxies with any precision.
However, following the principles of filter selection that were
detailed in \S\ref{source_of_signal}, it should be possible to
determine accurate photometric redshifts of $z$\lap 1.5 galaxies in
SIRTF surveys by augmenting their IRAC fluxes with $H$- and $K$-band
data.  Of course, if optical data are available --- as, for example,
will be the case for the SIRTF GOODS survey --- it will be possible to
also use ``traditional'' (4000\AA-break-dependent) photometric
redshifts to estimate reliable distances to $z$\lap 1 SIRTF-selected
objects.

\item In addition to the large scatter within 0$\leq$$z$\lap1.5, the
contours in the top two rows of Fig.~\ref{simulate_auto.fig} show that
$z$\lap 1.5 galaxies can be easily mistaken for very young (age \lap
10Myr) stellar populations at nearly {\it any} redshift.  Conversely,
extremely young stellar populations at high redshift ($z$\gap 1.5) can
easily be misidentified as $z$\lap 1.5 galaxies.  This degeneracy is a
result of the similarity between the colors of low-$z$ objects and
those of very young stellar populations at any redshift
(Fig.~\ref{IRAC_colours.fig}).  Because of this degeneracy, SIRTF data
alone cannot be used as a reliable way to say that a galaxy is at
$z$\lap1.5; neither can they be used to unambiguously identify
high-$z$ galaxies that are dominated by very young stellar
populations.  Such degeneracies could possibly be broken with the
addition of UV--optical data that enable the use of the Lyman break
technique (e.g.\ Steidel et al.\ 1996) or of ``traditional''
photometric redshifts (e.g.\ Sawicki et al.\ 1997).

\item Photometric redshift of galaxies whose fits show them to be both
at $z$\gap 1.5 and older than $\sim$20 Myr {\it can} be relied on:
there exists no confusion in Fig~\ref{simulate_auto.fig} between
$z$\lap1.5 galaxies and older higher-$z$ objects, at least under the
assumption (in force here) that all the errors in photometric
redshifts arise from random errors in photometry.  Rejecting galaxies
whose photometric redshifts claim them to be either at $z$\lap1.5, or
younger than $\sim$20 Myr, leaves a reliable core of objects which
should be free of catastrophic errors in photometric redshift.

\item The expected scatter in photometric redshifts of objects with
$z$\gap1.5 and ages\gap20Myr is smallest at $z$$\approx$2 and
increases to higher redshifts.  This increase results because IRAC
colors become progressively flatter and more degenerate with
increasing redshift (see Fig.~\ref{IRAC_colours.fig}) as the \bump\
encroaches into the reddest (anchor) bandpass.  The addition of a
redder filter would provide an anchor bandpass that would improve the
accuracy of photometric redshifts at $z$$\approx$3 and beyond.
Fortunately, with IRAC data alone, the scatter is small at
$z$$\approx$2, which is precisely the epoch where spectroscopic
redshifts are hardest to secure.  The epoch around $z$$\approx$2 is
also the era where the cosmic star formation appears to plateau (e.g.,
Sawicki, Lin, \& Yee 1997; Steidel et al.\ 1999) after its strong rise
with lookback time from $z$=0 to $z$$\approx$1 (Lilly et al.\ 1996);
the ability of SIRTF data to reliably select $z$$\approx$2 galaxies
makes SIRTF surveys well-poised to fill the gap in our picture of the
galaxy population at this crucial transitional epoch.

\item IRAC data alone are insufficient to provide strong constraints
on galaxy ages.  The ages of intrinsically very young stellar
populations (age\lap 10Myr) could in principle be constrained well if
one could break the strong degeneracy between such objects and those
at $z$\lap 1.5 by the addition of bluer filters, or --- ideally ---
spectroscopic redshifts.  The ages of intrinsically older galaxies
(age\gap 50 Myr) are more difficult to constrain because IRAC colors at
high $z$ do not evolve quickly with time.  However, while the scatter
in photometric redshifts increases with redshift past $z$$\approx$2,
the accuracy of age determination improves because of the divergence
of (primarily) \iracL\ colors at higher redshifts; the addition of
bluer IR filters should allow ages to be constrain even more
precisely.

\end{enumerate}
	
Thus, it appears that SIRTF IRAC colors alone may be sufficient to
measure photometric redshifts of high-$z$ galaxies, provided that care
is taken to reject objects whose fits claim them to be either at
$z$\lap 1.5 or younger than $\sim$20 Myr (these redshift and age
cut-offs can be adjusted depending on the size of the photometric
uncertainties and on the desired balance between completeness and
contamination of the sample).  At $z$\lap 1.5 the accuracy of
photometric redshifts would greatly improve with the addition of $H$
and $K$ fluxes, while at higher redshifts ($z$\gap 2.5) low-$z$
contaminants can potentially be separated from young high-$z$ galaxies
by the addition of UV-optical data that are sensitive to the presence
of the Lyman break.  Overall, random photometric errors of reasonable
size appear to allow photometric redshifts to be determined using
SIRTF data over selected regions of parameter space.

\subsection{Systematic Errors}\label{systematics}

So far only the effects of random photometric errors have been
considered.  However, given our incomplete state of knowledge about
the nature of high-$z$ galaxies, there exist many potential sources of
systematic error.  If the spectral energy distributions of high-$z$
galaxies are different from what we assume, our photometric estimates
of these galaxies' redshifts may be biased.  Metallicity effects,
deviations from the expected makeup of the stellar populations, and
the presence of dust can all modify the shape of the spectrum around
1.6\um, and thus can systematically bias photometric redshifts.  This
section considers what happens when the dust-free, constantly
star-forming, $Z$=0.2\Zsun\ template grid of \S\ref{photometricerrors}
is used to fit ``observed'' galaxies of discrepant metallicity, star
formation history, or dust content.

\subsubsection{Dust}

High-$z$ galaxies can be shrouded by substantial amounts of
interstellar dust.  For example, Sawicki \& Yee (1998) found that some
Lyman Break Galaxies in the HDF may be reddened by as much as
$E(B-V)$$\approx$0.4, or $A_V$$\approx$2 (see also Papovich,
Dickinson, \& Ferguson 2001; Shapley et al.\ 2001).  Because
interstellar dust tends to preferentially suppress flux in the blue,
it can be expected to both shift the apparent wavelength of the \bump\
and to make the stellar population appear older.

\begin{figure}
\plotone{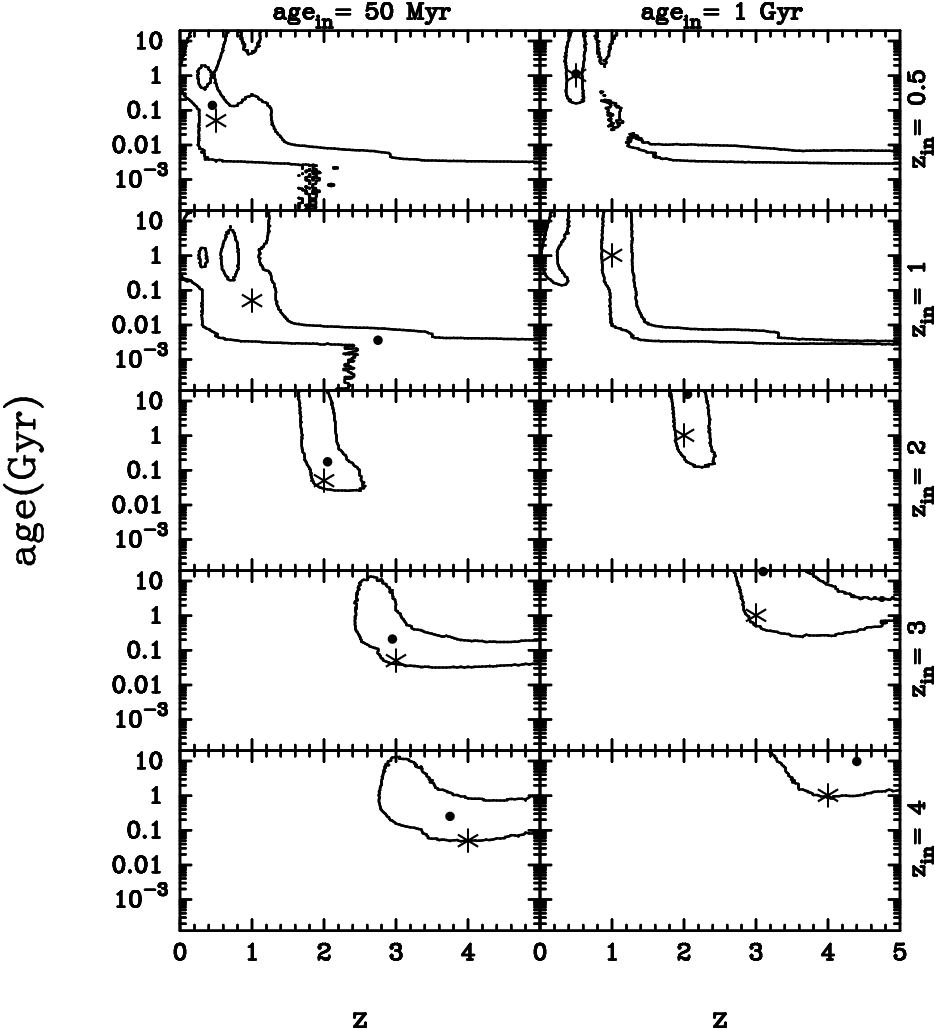} 
\caption{\label{simulate_dust.fig} The effect of mismatch between the
amounts of dust in the fitted galaxy and the fitting grid of
templates.  Asterisks show the input redshifts and ages (\zin,
\agein), while filled circles indicate the best fit solutions (\zfit,
\agefit).  The ``observed'' objects are enshrouded in substantial
amounts of dust (\ebv=0.3), while the fitting templates are dust-free.
The close match between \zin\ and \zfit\ values in this Figure, as
well as the similarity in the shape of confidence regions between this
Figure and Fig.~\ref{simulate_auto.fig}, indicates that the recovery
of redshift information is not strongly affected by a mismatch in dust
obscuration between ``observed'' galaxies and fitting templates.  }
\end{figure}

To study the effect that dust may have on the accuracy of photometric
redshifts, templates constructed from constantly star-forming,
$Z$=0.2\Zsun, dust-free GISSEL models (i.e., identical to those in
\S\ref{photometricerrors}) were used to fit ``observed'' galaxies that
had the same metallicity and star formation history as these fitting
templates, but were reddened with \ebv=0.3 following the Calzetti
(1997) starburst dust curve.  Fit results are shown in
Fig.~\ref{simulate_dust.fig}; as with Fig.~\ref{simulate_auto.fig},
input galaxies of \agein=50 Myr and 1Gyr, and located at \zin=0.5, 1,
2, 3, and 4, were used.  The asterisks in Fig~\ref{simulate_dust.fig}
indicate the locations of these input parameters, while --- as in
Fig.~\ref{simulate_auto.fig} --- filled circles and contours show the
best-fit values and regions of 95\% confidence for $\sigma(i)$=(0.05,
0.05, 0.075, 0.1) Gaussian photometric errors.

The dust-free template grid recovers faithfully (certainly to well
within the 95\% error bars in \zfit) the redshifts of the dusty
``observed'' galaxies.  At $z$$>$3, the recovered {\it ages} are
noticeably older than the input ones --- presumably because dust
preferentially suppresses flux blueward of the \bump, thereby
mimicking the redder SEDs of older stellar populations.  Nevertheless,
the ability to determine redshifts from IRAC photometry is not
compromised by a mismatch in dust content between the target galaxy
and the fitting template grid.  Provided the earlier caveats against
accepting fits with $z_{fit}$\lap 1.5 or age$_{fit}$\lap 20 Gyr are
heeded, IRAC colors should allow faithful measurements of redshifts
even in the presence of substantial amounts of dust.

\subsubsection{Star Formation History} 

\begin{figure}
\plotone{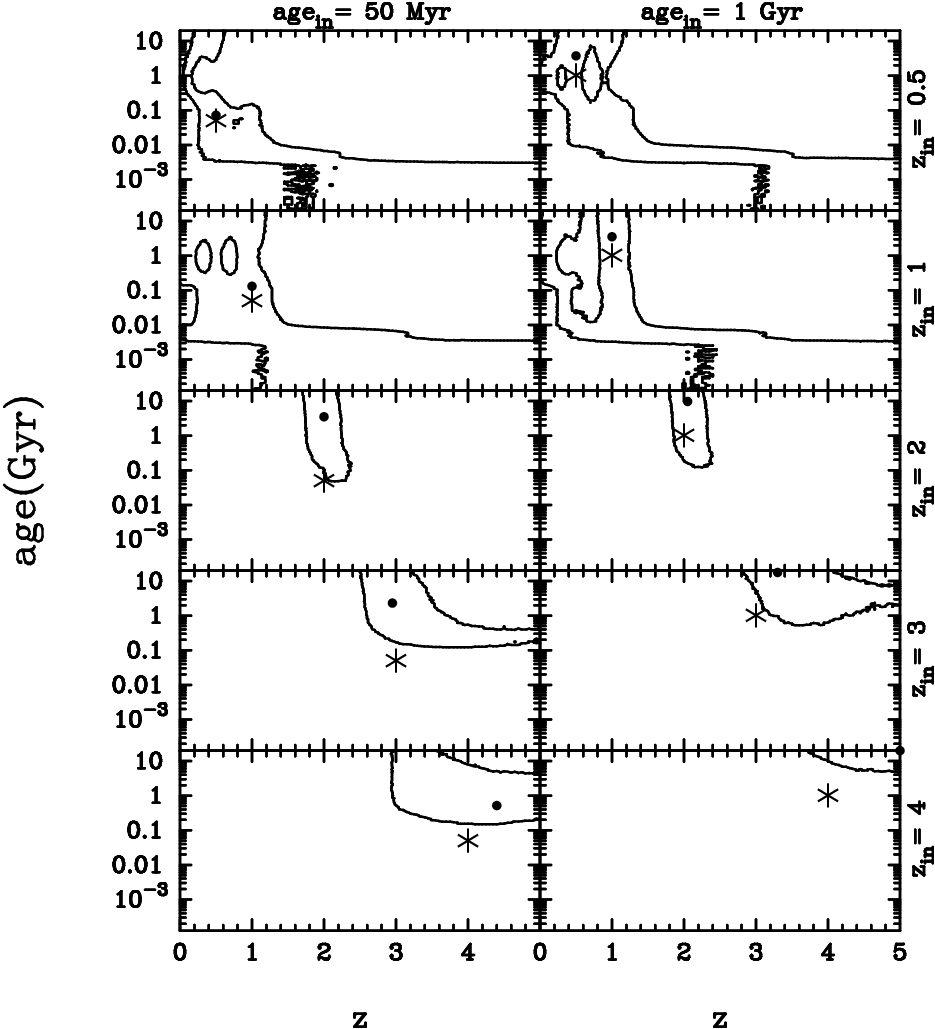} 
\caption{\label{simulate_SFhistory.fig} The effect of mismatch between
the star formation histories of the fitted galaxy and the fitting
templates.  The meaning of symbols and contours is the same as in
Figs.~\ref{simulate_auto.fig}~\&~\ref{simulate_dust.fig}.  The
``observed'' objects have been formed in instantaneous bursts of star
formation, while the fitting templates are made from models with
constant star formation.  The assumed star formation history does not
appear to have a strong effect on the ability to recover the correct
redshift.  }
\end{figure}

The star formation histories of high-$z$ galaxies are at present not
well understood (see, e.g., Sawicki \& Yee 1998; Papovich et al.\
2001; Shapley et al.\ 2001).  Two extreme illustrative scenarios are
(1) an instantaneous burst of star formation and (2) a constantly
star-forming system.  The effect of star formation history on the
fidelity of photometric redshifts was studied by using the standard
grid of constantly star-forming, $Z$=0.2\Zsun, dust-free templates to
fit ``observed'' galaxies that have the same metallicity and dust
content as the grid, but which underwent only a single, instantaneous
burst of star formation that occurred either 50Myr or 1Gyr (left and
right-hand panels in Fig.~\ref{simulate_SFhistory.fig}, respectively)
before being ``observed.''

As Fig.~\ref{simulate_SFhistory.fig} shows, despite a mismatch in star
formation history between ``observed'' objects and fitting templates,
redshifts of \zin$>$1.5 galaxies are recovered faithfully.  At
\zin$>$3, the ages of galaxies tend to be overestimated --- a result
of the deficit of young, blue stars in the instantaneous-burst
``observed'' galaxies that makes them seem older than they are when
fit with constant SFR templates.  Nevertheless, IRAC-based photometric
redshifts can be trusted provided that, as before, one avoids objects
with $z_{fit}$\lap 1.5 or age$_{fit}$\lap 20Gyr.

\subsubsection{Metallicity}

The creation of the H$^-$ ion relies on the presence of a supply of
free electrons, and so the abundance of metals in stellar atmospheres
may affect the strength of the \bump.  A metal-poor object can be
expected to have a less pronounced \bump\ and a modified shape of the
SED either side of the bump, potentially affecting both the recovered
redshift and age of the target galaxy.  Metallicities of high-$z$
galaxies are as yet poorly constrained, ranging at least over
$Z$=0.2--1\Zsun\ for Lyman Break Galaxies (Kobulnicky \& Koo 2000;
Pettini et al.\ 2000, 2001; Teplitz et al.\ 2000), and extending to
even lower values for damped Ly$\alpha$ systems (e.g., Pettini et al.\
1999; Prochaska, Gawiser, \& Wolfe 2001).  To study the effect that
metallicity has on the fidelity of photometric redshifts, the standard
grid of constantly star-forming, $Z$=0.2\Zsun, dust-free templates was
used to fit ``observed'' galaxies that have the same star formation
history and dust content as the grid, but which have either higher
($Z_{in}$ =\Zsun), or lower ($Z_{in}$=0.02\Zsun) metallicities.

\begin{figure}
\plotone{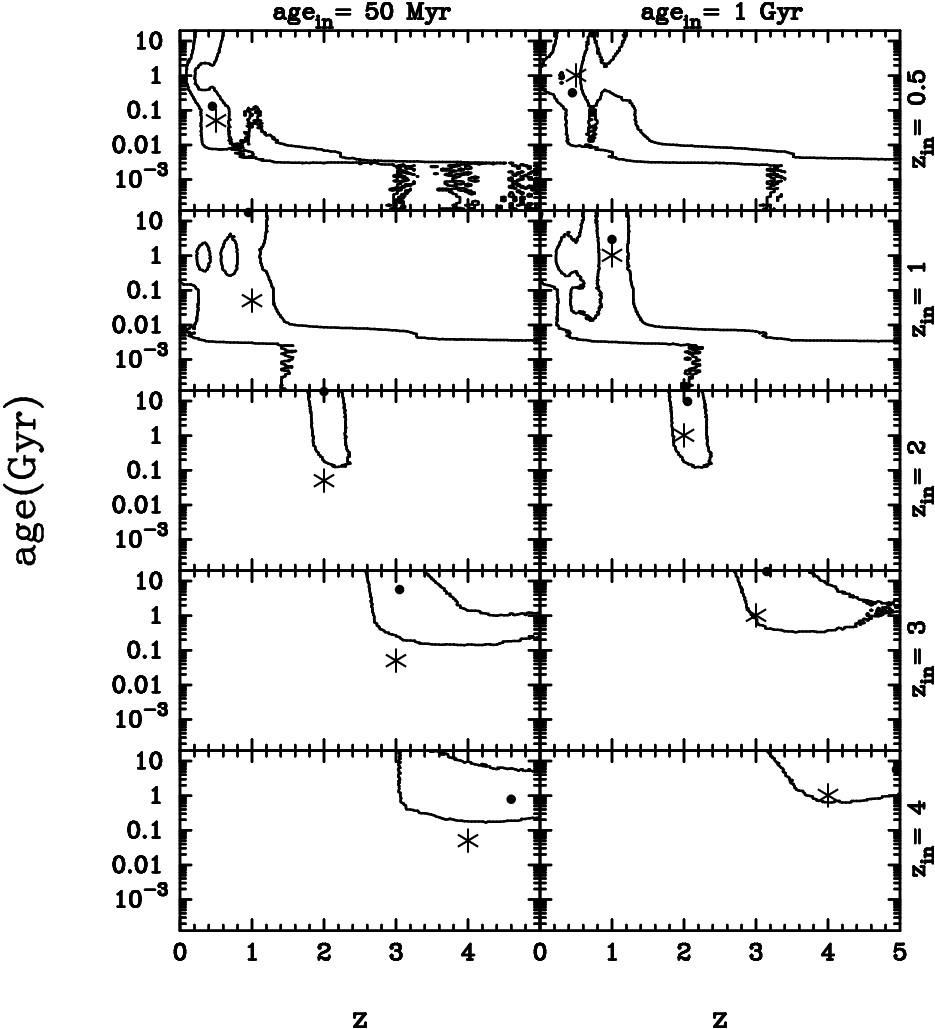}
\caption{\label{simulate_metallicity1.0Zsun.fig} The effect of
mismatch in the metallicities of the object and the fitting grid of
templates.  The meaning of symbols and contours is the same as in
Figs.~\ref{simulate_auto.fig}--\ref{simulate_SFhistory.fig}.  The
``observed'' object has solar metallicity, while the fitting grid has
$Z$=0.2\Zsun.}
\end{figure}

\begin{figure}
\plotone{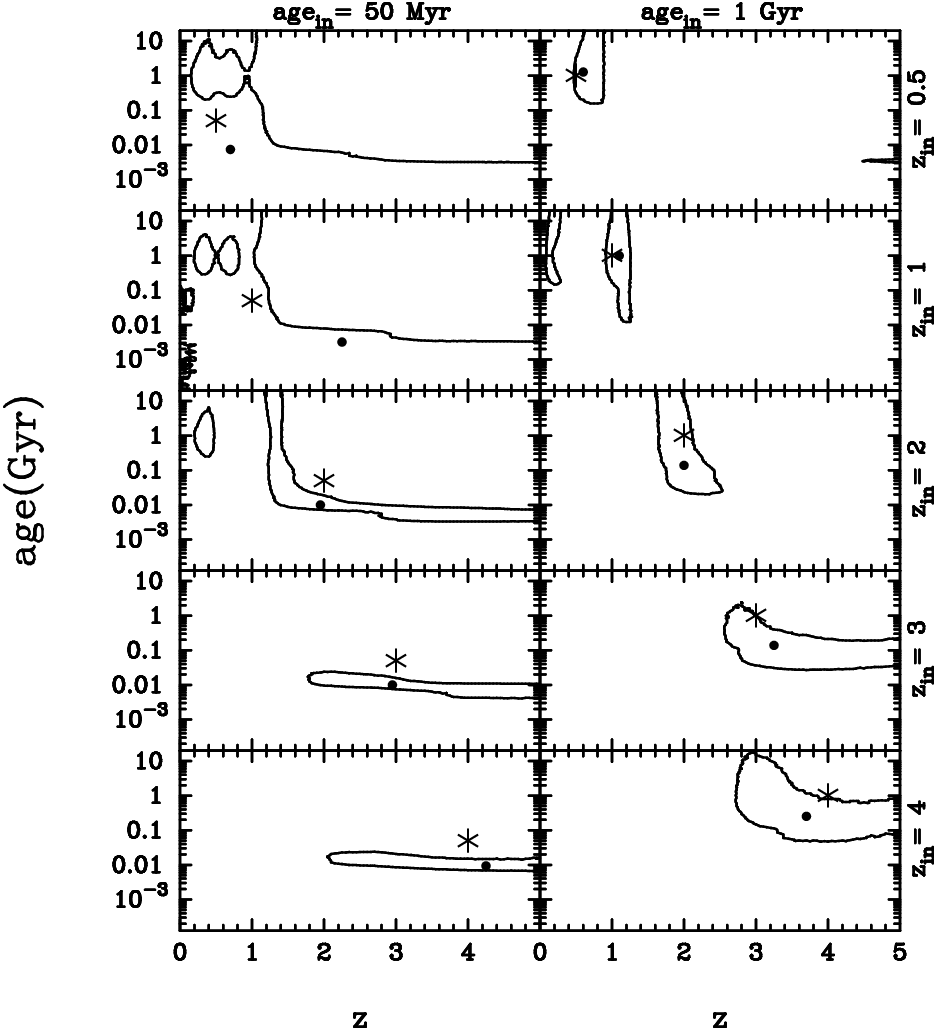}
\caption{\label{simulate_metallicity0.02Zsun.fig} The effect of
mismatch in the metallicities of the object and the fitting grid of
templates.  The meaning of symbols and contours is the same as in
Figs.~\ref{simulate_auto.fig}--\ref{simulate_metallicity1.0Zsun.fig}.
The ``observed'' object has a metallicity of 0.02\Zsun, while the
fitting template grid has $Z$=0.2\Zsun.  }
\end{figure}

Fig.~\ref{simulate_metallicity1.0Zsun.fig} shows the results of the
fit for ``observed'' galaxies of solar metallicity
(Fig.~\ref{simulate_metallicity0.02Zsun.fig} shows fit results for
$Z_{in}$=0.02\Zsun).  As Fig.~\ref{simulate_metallicity1.0Zsun.fig}
illustrates, photometric redshifts of ``observed'' galaxies that are
more metal-rich than the fitting template grid are not strongly
affected.  At high redshift (\zin$\approx$4) the ages derived from the
fit are overestimated --- as expected given that metal-rich galaxies
have a stronger \bump\ that makes their age-sensitive
$\lambda_{rest}$$<$1.6\um\ SEDs appear more depressed and, hence,
older.

Fit results for ``observed'' galaxies that are more metal-poor than
the template grid are shown in
Fig.~\ref{simulate_metallicity0.02Zsun.fig}.  Not surprisingly, at
\zin\gap 3 fit ages are younger than the true ages, \agein. For
intrinsically old galaxies (\agein=1Gyr) this does not affect the
recovery of redshift significantly.  However, for intrinsically young
objects (\agein=50Myr), this underestimate of the true age pushes the
fit into the region at \agefit$\approx$10Myr that can be heavily
degenerate with \zfit\lap1.5 solutions.  Thus, while the redshifts of
metal-rich high-$z$ galaxies should be recovered faithfully, IRAC
colors alone are not sufficient to unambiguously identify high-$z$,
very metal-poor galaxies that are dominated by young stellar
populations (see also Simpson \& Eisenhardt, 1999).

\section{SUMMARY AND DISCUSSION}\label{summary}

This paper considers the viability of the \bump\ as a photometric
redshift indicator.  In \S\ref{HDF} it was shown that $JKHL$
photometry can be used to estimate the redshifts of $z$$<$1 galaxies,
thus demonstrating with real data that the \bump\ is a viable
photometric redshift indicator.  The analysis of the dependence of
galaxy colors on redshift (\S\ref{source_of_signal}) leads to the
conclusion that photometric redshift signal at a given redshift comes
primarily from two filters: an ``anchor'' filter which at all
redshifts of interest remains in the invariant part of the spectrum
redward of the \bump, and a second filter whose redshifted bandpass
corresponds to 1.6\um.  To cover an extensive redshift range and to
reduce redshift degeneracies, one needs --- in addition to the
long-wavelength anchor filter that is always redward of (1+$z$)1.6\um\
--- multiple filters that span the possible range of redshifted
1.6\um.  Filters that are located blueward of the redshifted \bump\ do
not constrain redshift well, but instead provide information about the
age of the stellar population.  In this manner, the $HKL$ filter set
was used to determine redshifts of galaxies at $z$$\approx$0--1 (the
$J$ filter does not provide redshift information but could be used as
an age indicator), while the SIRTF IRAC \iracfilters\ filters can be
expected to work at higher redshifts.

Simulations (\S\ref{photometricerrors}) suggest that SIRTF IRAC colors
alone may be sufficient to measure photometric redshifts of high-$z$
galaxies, provided that care is taken to reject the degenerate objects
whose fits claim them to be either at $z$\lap 1.5 or younger than
$\sim$20 Myr (these redshift and age cut-offs can be adjusted
depending on the levels of completeness and contamination that are
acceptable).  At higher redshifts ($z$\gap 2.5) low-$z$ contaminants
can be separated from young high-$z$ galaxies by the addition of
UV-optical data that are sensitive to the presence of the Lyman break;
at $z$\lap 1.5, the accuracy of photometric redshifts would be
restored with the addition of $H$ and $K$ bands.
 
As was illustrated in \S\ref{systematics}, the ability to determine
photometric redshifts from IRAC photometry should not be significantly
compromised by the unknown metallicities, dust content, and star
formation histories of high-$z$ galaxies.  Provided the caveats about
rejecting galaxies with \zfit \lap 1.5 or \agefit \lap 20Myr are
followed, SIRTF-based photometric redshifts should provide a powerful
way to estimate the redshifts of distant galaxies.  With SIRTF IRAC
data alone, photometric redshifts can be expected to be most accurate
for galaxies at $z$$\approx$2, with the accuracy decreasing to higher
redshifts; they can be expected to be completely inaccurate for
galaxies at $z$\lap 1.5.  The epoch at $z$$\approx$2 is the
transitional era where the cosmic star formation appears to plateau
(e.g., Sawicki, Lin, \& Yee 1997; Steidel et al.\ 1999) after its
rapid rise with lookback time over 0$<$$z$\lap 1 (Lilly et al.\ 1996);
the expected high accuracy of IRAC photometric redshifts at
$z$$\approx$2 means that SIRTF surveys are well-poised to study
galaxies at this hitherto unexplored epoch of transition.

Although the \bump\ can be used to provide a good estimate of redshift
from SIRTF data, particularly at $z$$\approx$2, ages of stellar
populations cannot be determined with great precision from SIRTF data
alone at any redshift: large scatter in \agefit\ can be expected in
the presence of even small photometric uncertainties
(\S\ref{photometricerrors}), while significant systematic effects will
be present due to the unknown metallicity, star formation history, and
dust content of the target galaxy (\S\ref{systematics}).  The addition
of optical and near-IR data could alleviate some of these problems:
while SIRTF data are used to estimate redshifts, near-IR data that
span the age-sensitive rest-frame 4000\AA\ break can be added to
estimate ages, and dust effects can be constrained in the rest-UV with
optical data (see, e.g., Sawicki \& Yee, 1998).  The addition of
spectroscopic redshifts might then free the SIRTF photometry to be
used to better constrain star formation histories and/or metallicities
of distant galaxies.  It is beyond the scope of this paper to explore
what constraints can be placed on the evolutionary parameters of
high-$z$ galaxies by multiwavelength photometry augmented by
spectroscopic redshifts; however, such studies that combine
spectroscopy with photometry that ranges from rest-frame UV to IR are
sure to be an important path to understanding the nature of distant
galaxies.

Although \S\ref{HDF} illustrates that reliable photometric redshifts
are possible using the \bump, the essential test of the \bump\ as a
photometric redshift indicator for SIRTF will come only in the form of
a comparison between spectroscopic redshifts and photometric ones
based on SIRTF data.  The \bump\ technique is expected to be most
useful at $z$$\approx$2 --- precisely where our ability to determine
spectroscopic redshifts is limited by the lack of readily accessible
strong spectral features.  However, we should be able to reliably
assemble large samples of $z$$\approx$2 galaxies by first using
photometric redshifts to select $z$$\approx$2 candidates, then
obtaining spectroscopic redshifts using deep spectroscopy that targets
absorption lines that are weak but accessible in the optical (e.g.,
\ion{C}{4} 1550, \ion{Mg}{2} 1670), and finally using the
spectroscopic redshifts to test and calibrate the photometric
redshifts.  The \bump\ photometric redshift technique applied to SIRTF
(and, later, NGST) data appears to be ideally suited to allow a
detailed study of galaxies at the crucial epoch around $z$$\approx$2.

\acknowledgements This work has benefitted from my interactions with
Doug Johnstone, Keith Matthews, Gerry Neugebauer, Tom Soifer, Howard
Yee, and the anonymous referee.  I thank all of them for their diverse
contributions.


\clearpage



\twocolumn

\clearpage

\begin{deluxetable}{llcccccc}
\tablewidth{0pt} \tablecaption{\label{photometry.tab}Photometry and redshifts of HDF galaxies detected in $L$-band } 
\tablehead{ \colhead{R.A.\tablenotemark{a}}& \colhead{Dec.\tablenotemark{a}} & \colhead{$J_{AB}$\tablenotemark{b}} & \colhead{$H_{AB}$\tablenotemark{b}} & \colhead{$K_{AB}$\tablenotemark{b}} & \colhead{$L_{3.2\mu m,AB}$\tablenotemark{b,c}} & \colhead{$z_{spec}$} & \colhead{$z_{phot}$}}
\startdata
12 36 44.017 & +62 12 50.11 & 20.71$\pm$0.09 & 20.33$\pm$0.06 & 20.03$\pm$0.06 & 20.04$\pm$0.14 & 0.557 & 0.70 \\
12 36 44.626 & +62 13 04.29 & 20.78$\pm$0.09 & 20.34$\pm$0.06 & 20.10$\pm$0.06 & 20.24$\pm$0.13 & 0.485 & 0.55 \\
12 36 48.088 & +62 13 09.21 & 20.10$\pm$0.07 & 19.76$\pm$0.05 & 19.56$\pm$0.05 & 19.74$\pm$0.09 & 0.476 & 0.50 \\
12 36 49.435 & +62 13 46.92 & 18.09$\pm$0.03 & 17.90$\pm$0.02 & 17.99$\pm$0.02 & 18.77$\pm$0.04 & 0.089 & 0.15 \\
12 36 49.541 & +62 14 06.85 & 21.34$\pm$0.12 & 20.96$\pm$0.09 & 20.59$\pm$0.07 & 20.55$\pm$0.17 & 0.752 & 0.80 \\
12 36 51.783 & +62 13 53.85 & 20.82$\pm$0.10 & 20.36$\pm$0.06 & 20.06$\pm$0.06 & 20.27$\pm$0.14 & 0.557 & 0.55 \\
12 36 53.916 & +62 12 54.26 & 20.58$\pm$0.09 & 20.17$\pm$0.06 & 19.87$\pm$0.05 & 20.10$\pm$0.10 & 0.642 & 0.55 \\
12 36 55.460 & +62 13 11.63 & 21.10$\pm$0.11 & 20.65$\pm$0.07 & 20.26$\pm$0.06 & 20.27$\pm$0.17 & 0.968 & 0.80 \\
12 36 56.675 & +62 12 45.51 & 20.13$\pm$0.07 & 19.75$\pm$0.05 & 19.52$\pm$0.05 & 19.80$\pm$0.08 & 0.518 & 0.45 \\
\enddata
\tablenotetext{a}{From Hogg et al.\ (2000).}
\tablenotetext{b}{Magnitudes within 2\arcsec diameter circular apertures. }
\tablenotetext{c}{Based on ``total'' magnitudes of Hogg et al.\ (2000) and converted from Vega to AB normalization assuming $L_{3.2\mu m,AB}$=$L_{3.2\mu m,Vega}$+2.77.}
\end{deluxetable}



\newpage

\begin{thebibliography}{}

\bibitem[Arnouts et al.\ (1999)]{arn99} Arnouts, S., Cristiani, S.,
Moscardini, L., Matarrese, S., Lucchin, F., Fontana, A., \& Giallongo,
E. 1999, \mnras, 540, 556

\bibitem[Bertin \& Arnouts (1996)]{ber96} Bertin, E. \& Arnouts,
S. 1996, \aaps, 117, 393

\bibitem[Bruzual \& Charlot (1993)]{bru93} Bruzual A., G., \& Charlot,
S.  1993, \apj, 405, 538

\bibitem[Calzetti (1997)]{cal97} Calzetti D. 1997, in AIP
Conf. Proc. 408, The Ultraviolet Universe at Low and High Redshift,
eds. W. H. Waller et al.\ (New York: AIP), 403

\bibitem[Cohen et al.\ (1996))]{coh96} Cohen, J.G., Cowie, L.L., Hogg,
D.W., Songaila, A., Blanford, R., Hu, E.M., \& Shopbell, P., 1996,
\apjl, 471, L5

\bibitem[Cohen et al.\ (1999)]{coh99} Cohen, J.G., Hogg, D.W., Blanford,
R., Cowie, L.L., Hu, E.M., Songaila, A., Shopbell, P., \& Richberg, K.,
2000, \apj, 538, 29

\bibitem[Connolly et al.\ (1995)]{con95} Connolly, A.J., Csabai, I.,
Szalay, A.S., Koo, D.C., Kron, R.G., \& Munn, J.A., 1995, \aj, 110,
2655

\bibitem[Dickinson et al.\ (1998)]{dic98} Dickinson, M., 1998, in {\it
The Hubble Deep Field,} eds.\ M.\ Livio, S.\ M.\ Fall, \& P.\ Madau
(Cambridge), 219

\bibitem[Fern\'andez-Soto et al.\ (1999)]{fer99} Fern\'andez-Soto, A.,
Lanzetta, K.M., \& Yahil, A. 1999, \apj, 513, 34

\bibitem[Gwyn \& Hartwick (1996)]{gwy96} Gwyn, S.D.J, \& Hartwick,
F.D.A. 1996, \apjl, 468, L77

\bibitem[Hogg et al.\ (2000)]{hog00} Hogg, D.W., Neugebauer, G., Cohen, J.G.,
 Dickinson, M., Djorgovski, S.G., Matthews, K., \& Soifer, B.T. 2000, 
\aj, 119, 1519

\bibitem[John (1988)]{joh88} John, T.L. 1988, \aap, 193, 189

\bibitem[Kobulnicky \& Koo (2000)]{kob00} Kobulnicky, H.A., \& Koo,
D.C., 2000, \apj, 542, 712

\bibitem[Lilly et al. (1996)]{lil96} Lilly, S.J., Le F\'evre, O.,
Hammer, F., \& Crampton, D. 1996, \apjl, 460, L1

\bibitem[Lowenthal et al.\ (1997)]{low97} Lowenthal, J.D., Koo, D.C., 
Guzm\'an, R., Gallego, J., Phillips, A.C., Vogt, N.P., Illingworth, G.D.,
\& Gronwall, C. 1997, \apj, 481, 673

\bibitem[Papovich et al.\ (2001)]{pap01} Papovich, C., Dickinson, M.,
Ferguson, H.C. 2001, \apj, 559, 620

\bibitem[Pettini et al 1997]{pet97} Pettini, M., Ellison, S. L.,
Steidel, C. C., Bowen, D. V.  1999, \apj, 510, 576

\bibitem[Pettini et al.\ (2001)]{pet01a} Pettini, M., Steidel, C.C.,
Adelberger, K.L., Dickinson, M., \& Giavalisco, M. 2000, \apj, 528, 96

\bibitem[Pettini et al.\ (2001)]{pet01b} Pettini, M., Shapley, A.E.,
Steidel, C.C., Cuby, J.-G., Dickinson, M., Moorwood, A.F.M.,
Adelberger, K.L., \& Giavalisco, M. 2001, \apj, 554, 981

\bibitem[Prochaska, Gawiser \& Wolfe (2001)]{pro01} Prochaska, J. X.,
Gawiser, E., \& Wolfe, A. M., 2001, ApJ, 552, 99

\bibitem[Rudnick et al.\ (2001)]{rud01} Rudnick, G., Franx, M., Rix,
H.-W., Moorwood, A., Kuijken, K., van Starkenburg, L..  van der Werf,
P., R\"ottgering, H., van Dokkum, P., Labb\'e I. 2001, \aj, 122, 2205

\bibitem[Sawicki et al.\ (1997)]{saw97} Sawicki, M.J., Lin, H., \& Yee,
H.K.C. 1997, \aj, 113, 1

\bibitem[Sawicki et al.\ (1998)]{saw98} Sawicki, M. \& Yee, H.K.C. 1998, 
\aj, 115, 1329

\bibitem[Shapley et al.\ (2001)]{sha01} Shapley, A.E., Steidel, C.C.,
Adelberger, K.L., Dickinson, M., Giavalisco, M., \& Pettini, M. 2001,
\apj, 562, 95

\bibitem[Simpson \& Eisenhardt (1999)]{sim99} Simpson, C. \& Eisenhardt,
P. 1999, \pasp, 111, 691

\bibitem[Steidel et al.\ (1996)]{ste96} Steidel, C.C., Giavalisco, M.,
Dickinson, M., \& Adelberger, K.L. 1996, \apj, 112, 352

\bibitem[Steidel et al.\ (1999)]{ste99} Steidel, C.C., Adelberger,
K.L., Giavalisco, M., Dickinson, M., \& Pettini, M. 1999, \apj, 519, 1

\bibitem[Teplitz et al.\ (2000)]{tep00} Teplitz, H.I., McLean, I.S., 
Becklin, E.E., Figer, D.F., Gilbert, A.M., Graham, J.R., Larkin, J.E., 
Levenson, N.A., \& Wilcox, M.K. 2000, \apj, 533, L65

\bibitem[Williams et al.\ (1996)]{wil96} Williams, R. E., Blacker, B., 
Dickinson, M., Van Dyke Dixon, W., Ferguson, H. C., Fruchter, A. S., 
Giavalisco, M., Gilliland, R. L., Heyer, I., Katsanis, R., Levay, Z., 
Lucas, R. A., McElroy, D. B., Petro, L., Postman, M., Adorf, H.-M., 
\& Hook, R. N. 1996, \aj, 112, 1335

\bibitem[Wright et al. (1994)]{wri94} Wright, E.L., Eisenhardt, P., \&
Fazio, G. 1994, \baas, 26, 893 (astro-ph/9407055)

\end{thebibliography}
\end{document}